\documentclass[useAMS,usegraphicx]{mn2e}
\usepackage{psfig}
\title[SDSS J153259.96$-$003944.1]{Long-term optical photometric
monitoring  of the quasar SDSS J153259.96$-$003944.1
}
\author[C.\ S.\ Stalin \& R.\ Srianand]
{C.\ S.\ Stalin$^{1,2}$ and R.\ Srianand$^{1}$ \\
$^{1}$ IUCAA, Post Bag 4, Ganeshkhind, Pune 411 007, India \\
$^{2}$ Aryabhatta Research Institute of Observational Sciences (ARIES), Manora Peak, 
Nainital 263 129, India}

\date{Accepted 2005 xxxxx. Received 2005 xxxxx
           }

\pagerange{\pageref{firstpage}--\pageref{lastpage}}
\pubyear{2005}

\begin{document}

\maketitle

\label{firstpage}

\begin{abstract}
We report optical Cousins R and I band monitoring observations of the 
high redshift ({\it z} = 4.67) QSO SDSS J153259.96$-$003944.1 that 
does not show detectable emission lines in its optical spectrum.
We show this object varies with a 
maximum amplitude of  $\sim$0.4 mag during a year 
and three months of monitoring. Combined with 
two other epochs of photometric data available in the literature, we show
the object has gradually faded by $\sim$0.9 mag during the period 
June 1998 $-$ April 2001. A linear least squares fit to all available 
observations gives a slope of $\sim$0.35 mag/yr which translates to  
$\sim$1.9 mag/yr in the rest frame of the quasar. 
Such a variability is higher than that typically seen in QSOs but consistent 
with that of BL Lacs, suggesting that the optical continuum is Doppler boosted. 
Alternatively, within photometric errors, the observed lightcurve is also 
consistent with the object 
going through a microlensing event. Photoionization model calculations
show the mass of the Broad Line Region to be few tens of $M_\odot$ similar 
to that of low luminosity Seyfert galaxies, but $\sim$2 orders of magnitude 
less than that of luminous quasars. Further frequent photometric/spectroscopic
monitoring is needed to support or refute the different alternatives
discussed here on the nature of SDSS J153259.96$-$003944.1.
 
\end{abstract}

\begin{keywords} quasars: general -- quasars: individual: SDSS J153259.96$-$003944.1 
\end{keywords}

\section{Introduction}

The diverse observational characteristics of Active Galactic Nuclei (AGN) have 
been reconciled in an unification scheme
(Antonucci 1993; Urry \& Padovani 1995). The basic idea is that all the AGNs 
have broad and narrow line emitting regions (BLR and NLR respectively) and
an obscuring torus.  Most of the diversity in the observed properties is caused
by differences in our viewing angles to the torus axis. It has also been found that some of the physical
parameters of AGNs show statistically significant correlations. For example, 
tight relationships exist between: (i) the radius of the BLR and the continuum 
luminosity (Corbett et al.\ 2003) and (ii) the black hole mass and the 
velocity dispersion of the host galaxy (Onken et al.\ 2004 and references therein).
With the advent of new very large surveys, some objects that appear
to depart from this standard AGN picture 
are beginning to emerge. Understanding these observations are important
to get a clearer picture of AGN formation and evolution.

\par
Several AGNs were recently found to have peculiar emission  line 
characteristics. In the Sloan Digital Sky Survey (SDSS), a few high redshift 
quasars have been discovered without emission 
lines (Anderson et al. 2001; Hall et al. 2004). 
%*** IT SHOULD BE
%MENTIONED WHY THEY ARE PICKED OUT AS QUASARS IN THE ABSENCE OF THE
%"DEFINING" BROAD EMISSION LINES ***
Among them the {\it z} = 4.67 quasar SDSS J153259.96$-$003944.1 
(hereafter referred to as SDSS J1533$-$00) first reported by  
Fan et al. (1999) is the object of interest in the present
study. The optical spectrum of this quasar is featureless redward of the
Ly$\alpha$ forest region; blueward of $\sim$6800 \AA, the spectrum has 
features due to Ly$\alpha$ absorption at {\it z} = 4.52.
Near-IR observations confirm the presence of a point source
and an extended nebulosity in SDSS J1533$-$00 (see Hutchings 2003).
Another object (2QZ J215454.3$-$305654) with similar characteristics
is also reported by Londish et al.\ (2004).

The weak (or vanishing)  emission line spectrum seen in SDSS J1533$-$00 
is typical of BL Lac 
objects, where the lines are presumably swamped by the beamed and boosted
continuum (Urry \& Padovani 1995). BL Lac objects are characterized by strong radio 
and X-ray emission, optical variability and strong (and variable) optical 
polarization due to synchrotron radiation from a relativistic jet.
Apart from the featureless spectrum, SDSS J1533$-00$ does not possess these
other characteristic properties of BL Lac objects. It was not detected in deep 
radio observations (3$\sigma$ upper limit of 60 $\mu$Jy) and was
not found to be optically polarized (with a 3$\sigma$ upper limit of 4\%; Fan et al.\ 1999).
It was found to be an extremely weak X-ray source and was 
not detected in pointed Chandra observations (Vignali et al.\ 2001). 

The absence of strong emission lines in the case of SDSS J1533$-$00 
can be due to one of the following reasons: (i) the optical continuum being 
Doppler boosted as in the case of BL Lacs; (ii) the optical continuum being 
amplified due to a gravitational microlensing event by a star in an
 intervening galaxy;
and (iii) absence of line emitting gas in the vicinity of the
central UV continuum source. As frequent photometry of the
source can confirm or reject the first two possibilities we have carried out 
photometric monitoring observations.  In section 2 
we outline the observational program, data reduction
procedure and the results of the photometric monitoring. Section 3 discusses
the nature of the source. Finally, our conclusions are given in Section 4. 
The cosmological model we consider in our study is 
$\Omega_m$ = 0.3, $\Omega_\Lambda$ = 0.7 and H$_0$ = 70 km s$^{-1}$ Mpc$^{-1}$.

%%%%%%%%%%%%
\section{Optical observations and Analysis} 
\subsection{Photometry}
Optical photometric observations in Cousins R and I band were carried out 
on 10 nights between January 2000 and April 2001 using 
the 104 cm Sampurnanand telescope of the Aryabhatta Research Institute of 
Observational Sciences (ARIES), Nainital. This is an RC system with a f/13 
beam (Sagar 1999). The detector used was a cryogenically cooled 
2048 $\times$ 2048 CCD mounted at the cassegrain focus. Each pixel of the 
CCD corresponds to 0.37 (arcsec)$^2$ and the entire CCD 
covers $13^{\prime} \times 13^{\prime}$ on the sky. Observations were done 
in $2 \times 2$ binned mode to improve S/N. Typical seeing was 
$\sim$2$^{\prime\prime}$ during most of our observations.
\begin{figure}
\hspace*{-0.5cm}\psfig{file=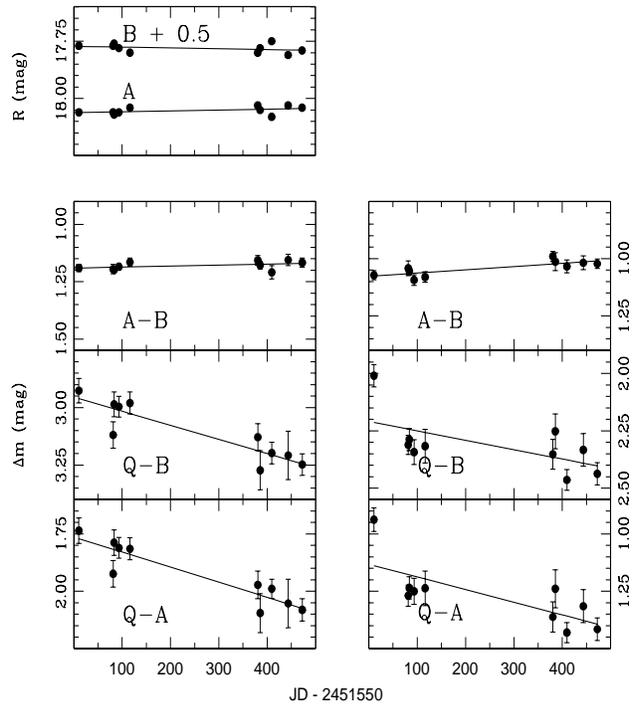,width=9cm,height=10cm}
\caption{Differential light curves (DLCs) of the quasar SDSS J1533$-$00 
with respect to two stars A (RA(2000) = 15:33:07.310; 
Dec(2000) = $-$00:39:04.0)
and B (RA(2000) = 15:33:04.110; Dec(2000) = $-$00:40:49.00).
The top most panel gives the standard R magnitudes of stars A and B.
The other panels give the DLCs of the pair of comparison stars 
and between quasar and the comparison stars as indicated within the panels. 
The left and right panels refer to R  and I bands respectively. The solid
lines are the linear least squares fits to the data.}
\label{fig1}
\end{figure}
%

%----------------------------------------------
%

\begin{table}
%\begin{center}
\caption{Log of observations and the results of photometry. Here, QSO$-$A, 
QSO$-$B and A$-$B refer to the differential instrumental magnitudes 
respectively between the quasar and the comparison stars A and B, and 
between the comparison stars themselves}
\begin{tabular}{llrccc} \hline
Date   &  $\lambda$  & Exp. &  QSO$-$A  & QSO$-$B & A$-$B  \\ 
       &             & Time &  (mag)    & (mag)   &  (mag) \\   
       &             & (sec)&           &         &      \\ \hline  
17.01.00  & R & 1200 &  1.74 $\pm$ 0.06 & 2.93 $\pm$ 0.05 & 1.19 $\pm$ 0.02 \\
          & I & 1800 &  0.94 $\pm$ 0.05 & 2.01 $\pm$ 0.05 & 1.07 $\pm$ 0.02 \\
28.03.00  & R & 1800 &  1.92 $\pm$ 0.06 & 3.12 $\pm$ 0.06 & 1.20 $\pm$ 0.02 \\
          & I & 1800 &  1.27 $\pm$ 0.05 & 2.31 $\pm$ 0.04 & 1.04 $\pm$ 0.03 \\
30.03.00  & R & 1800 &  1.79 $\pm$ 0.06 & 2.99 $\pm$ 0.05 & 1.20 $\pm$ 0.01 \\
          & I & 1800 &  1.24 $\pm$ 0.05 & 2.29 $\pm$ 0.05 & 1.05 $\pm$ 0.02 \\
09.04.00  & R & 1800 &  1.81 $\pm$ 0.05 & 3.00 $\pm$ 0.05 & 1.19 $\pm$ 0.01 \\
          & I & 1800 &  1.25 $\pm$ 0.06 & 2.34 $\pm$ 0.05 & 1.09 $\pm$ 0.02 \\
01.05.00  & R & 1800 &  1.82 $\pm$ 0.05 & 2.98 $\pm$ 0.05 & 1.17 $\pm$ 0.02 \\
          & I & 1800 &  1.24 $\pm$ 0.08 & 2.32 $\pm$ 0.07 & 1.08 $\pm$ 0.02 \\
21.01.01  & R & 1200 &  1.97 $\pm$ 0.06 & 3.13 $\pm$ 0.06 & 1.16 $\pm$ 0.02 \\
          & I & 1200 &  1.36 $\pm$ 0.07 & 2.35 $\pm$ 0.07 & 0.99 $\pm$ 0.02 \\
26.01.01  & R &  900 &  2.10 $\pm$ 0.09 & 3.27 $\pm$ 0.09 & 1.18 $\pm$ 0.02 \\
          & I &  900 &  1.24 $\pm$ 0.08 & 2.25 $\pm$ 0.08 & 1.01 $\pm$ 0.04 \\
19.02.01  & R & 1800 &  1.99 $\pm$ 0.04 & 3.20 $\pm$ 0.05 & 1.21 $\pm$ 0.03 \\
          & I & 1800 &  1.43 $\pm$ 0.04 & 2.46 $\pm$ 0.05 & 1.03 $\pm$ 0.03 \\
25.03.01  & R &  300 &  2.05 $\pm$ 0.11 & 3.21 $\pm$ 0.11 & 1.16 $\pm$ 0.02 \\
          & I &  300 &  1.32 $\pm$ 0.07 & 2.33 $\pm$ 0.07 & 1.02 $\pm$ 0.03 \\
23.04.01  & R & 1200 &  2.08 $\pm$ 0.05 & 3.25 $\pm$ 0.05 & 1.17 $\pm$ 0.02 \\
          & I & 1200 &  1.42 $\pm$ 0.05 & 2.44 $\pm$ 0.05 & 1.02 $\pm$ 0.02 \\ \hline
\end{tabular}
\end{table}

Initial processing of the images (bias
subtraction, flat fielding and cosmic ray removal) were
done using IRAF\footnote{The Image Reduction and Analysis Facility (IRAF) is
distributed by the National Astronomy Observatories, which is operated
by the Association of Universities for Research in Astronomy Inc. (AURA) under
cooperative agreement with the National Science Foundation} routines, whereas 
PSF fitting photometry was carried out using the routines in 
MIDAS\footnote{Munich Image Data Analysis Systems; trademark of the European 
Southern Observatory (ESO)}.  To look for variability, differential 
light curves (DLCs) of the quasar were generated  
with respect to stars A and B situated on the observed frames.
Typical error in our photometry is around 0.03 mag for the reference
stars and between 0.03 and 0.1 mag for the quasar.
The log of observations and the results of the photometry are
given in Table 1. The DLCs in R and I filters
with respect to stars A (RA(2000) = 15:33:07.310; 
Dec(2000) = $-$00:39:04.0) and B (RA(2000) = 15:33:04.110; 
Dec(2000) = $-$00:40:49.00) are shown in Fig. 1. From Fig. 1, 
it appears that the quasar shows a gradual fading during the 
period of our observations in both R and I bands, though of course 
more complex behavior cannot be excluded. In particular, as we have
gaps in our lightcurve, any non-linear fluctuations (such as flares) 
cannot be ruled out. Linear least squares
fits to the QSO DLCs in both R and I bands give a similar slopes of 
$\sim$0.20 mag/year. However, the QSO DLCs in I band show large
scatter compared to the R band data. This is due to the worse
quality of the I band data due to problems in flat fielding.
Also,  since the R band is near the maximum of the response 
curve of the CCD, we consider only the R band data in any further analysis. 
The amplitude of variability was calculated using 
$A_{max} = \sqrt{(D_{max} - D_{min})^2 - 2\sigma^2}$, 
where $D_{max}$ ($D_{min}$) are the maximum (minimum) in the quasar DLC 
and $\sigma$ the average error in the corresponding DLC (Romero et al. 1999). 
The object was found to show a peak to peak amplitude of variability ($A_{max}$)
of about 0.34 mag between January 2000 and April 2001. 

\begin{figure}
\hspace*{-0.5cm}\psfig{file=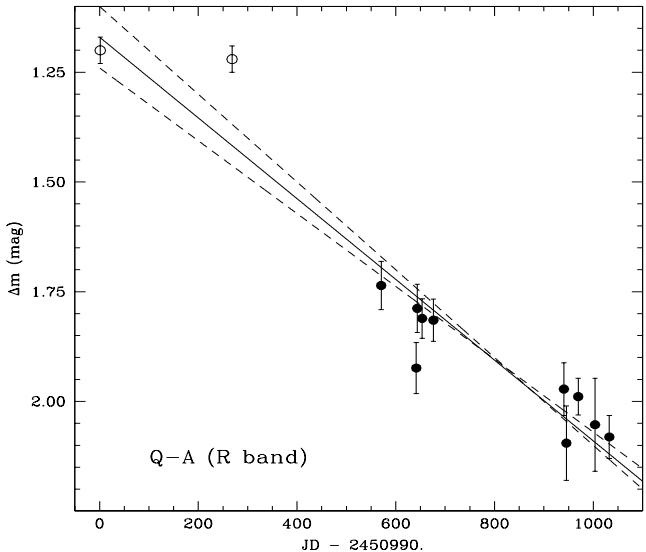,width=9cm,height=8cm}
\caption{Differential R band light curve of the quasar SDSS J1533$-$00 
with respect to star A (RA(2000) = 15:33:07.310; 
Dec(2000) = $-$00:39:04.0) including the observations of 
Fan et al. (1999) (shown as open circles).  Continuous and dotted lines give 
respectively the linear least squares fit and 1$\sigma$ error to the data.}
\label{fig2}
\end{figure}

SDSS J1533$-$00 was also observed for two epochs (27 June 1998 and
21 March 1999) by the SDSS team (Fan et al.\ 1999). No clear variation
is observed between these two epochs. Converting their $i^\ast$ magnitudes 
using equations (1) and (2)  given by Fukugita et al. (1996) and the 
zero point given by Bessel (1979), the R band magnitudes are estimated as 19.66 
and 19.68 on 27 June 1998 and 21 March 1999 respectively. In order to compare 
these quasar magnitudes to its magnitude during the epoch of our observations, 
we converted the quasar's instrumental magnitudes during our observations to 
apparent R magnitudes by the technique of differential photometry using Star A 
at the position (RA(2000) = 15:33:07.310; 
Dec(2000) = $-$00:39:04.0 and  whose R band magnitude of 18.46 mag is 
taken from USNO--B; Monet et al. 2003) available on the quasar CCD frame. 
The apparent R magnitude 
of the quasar thus calculated during our period of observations range from 
20.20 to 20.56 mag. Thus, combining our observations with those reported by 
the SDSS team, we find that the quasar has faded by $\sim$0.9 mag 
over a period of about 3  years from June 1998 to April 2001. Using a simple 
least squares fit to all the available data we get (see Fig.~2),
\begin{equation}
\Delta m(mag) = (1.17\pm0.07) + (0.34\pm0.03)\times t(yr).
\end{equation}
Here, $\Delta m$ is the differential magnitude of the quasar with respect
to the reference star A and $t$ is the time in the observer's frame.

\subsection{Spectroscopy}
\begin{figure*}
\vspace*{-0.9cm}
%\includegraphics{figure2.ps}}
%\hspace*{-0.2cm}\psfig{file=figure3.ps,width=17cm,height=8cm}
\hspace*{-1.5cm}\psfig{file=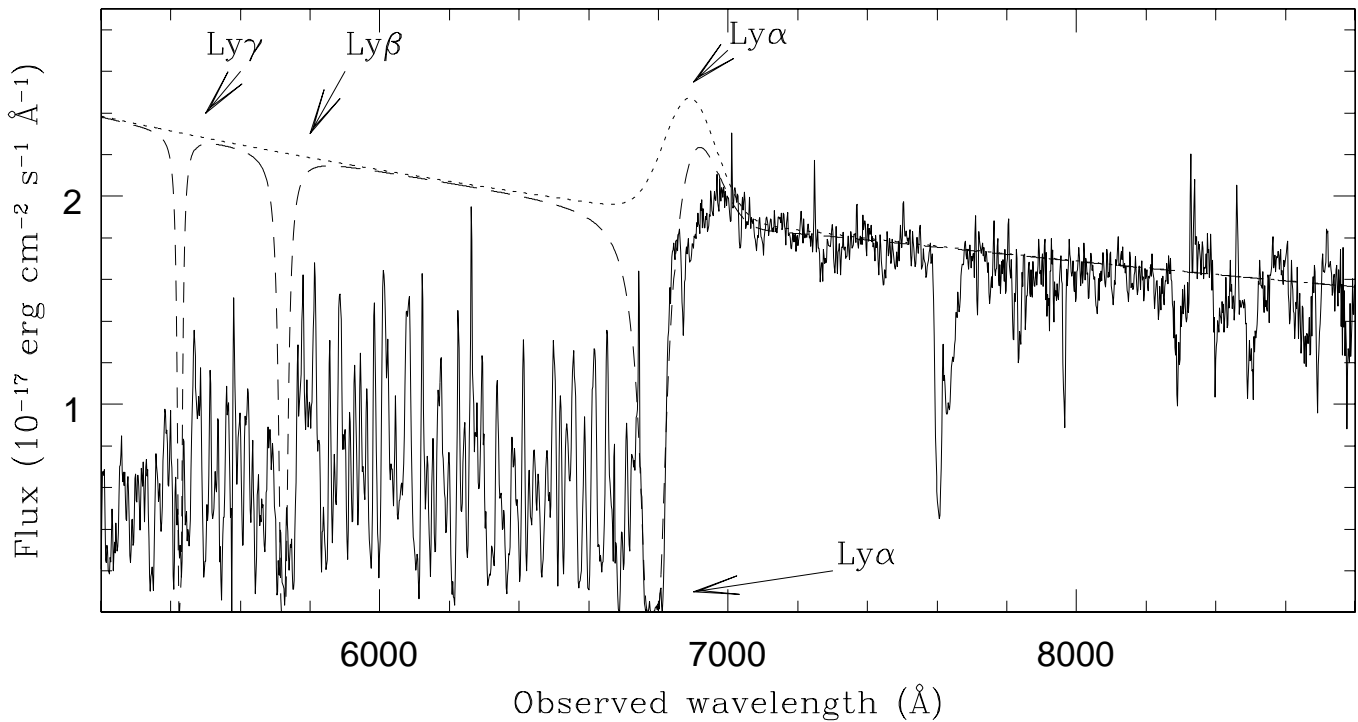}
\vspace*{-0.9cm}
\caption{The observed optical spectrum of the quasar using the LRIS 
spectrograph on Keck II (solid line) along with the fitted absorption and 
emission lines. The dotted line shows the modeled spectrum
containing only the Ly$\alpha$ emission line
whereas the dashed line shows the modeled spectrum containing
both Ly$\alpha$ emission and absorption lines of Ly$\alpha$, 
Ly$\beta$ and Ly$\gamma$.}
\label{fig3}
\end{figure*}

We use the optical spectrum of SDSS J1533$-$00, obtained by Fan et al.\
(1999) with the Keck II telescope using the Low Resolution Imaging Spectrograph,
to obtain the optical continuum spectral index, $\alpha_0$ (defined through,
$S_\nu\propto\nu^{\alpha_0}$), and an upper limit of the Ly${\alpha}$ 
emission line flux.  The observed spectrum is well fitted with the 
existence of Ly${\alpha}$ absorption system at {\it $z_{abs}$} = 4.58 with 
N(H{\sc i}) = 1.1 $\times 10^{21}$ cm$^{-2}$, Ly${\alpha}$ emission line with 
a FWHM of 105~\AA~ (or a velocity width of 4630 km/s) at an emission redshift of {\it $z_{em}$} = 4.67 and
$\alpha_0=-0.8$. N(H{\sc i}) in the absorbing gas cloud is obtained
by self-consistently fitting Ly$\alpha$, Ly$\beta$ and Ly$\gamma$ lines.
The fit to the spectrum is shown in Fig.~\ref{fig3}. Inclusion of 
Ly$\alpha$ emission in the fitting improves the fitting of the Ly$\alpha$
absorption line. However, as noticed by Fan et al.\ (1999) the higher Lyman 
series lines do not show complete absorption suggesting the gas producing the 
absorption cannot be a strong damped Ly$\alpha$ system. Thus our fits will
over predict the N(H{\sc i}) in the absorbing gas and
the derived upper limit of the Ly$\alpha$ emission line flux. 

The estimated flux of the fitted Ly$\alpha$ emission  line
is $\le1.06  \times 10^{-15}$ erg cm$^{-2}$ s$^{-1}$.  This corresponds
to an emitted luminosity of $\le2.2\times10^{44}$ erg s$^{-1}$. 
%for
%the cosmological model we consider in our study
%($\Omega_m$ = 0.3, $\Omega_\Lambda$ = 0.7 and H$_0$ = 70 km s$^{-1}$ Mpc$^{-1}$).
Using the fitted continuum and emission line flux we get an upper
limit ($\simeq$ 9.9~\AA) on the rest frame Ly$\alpha$ emission 
line equivalent width.

\subsubsection{Excess continuum flux}
As discussed before, the lack of emission lines in the spectrum of 
SDSS J1533$-$00 can be due to the continuum emission being enhanced
either by Doppler boosting of the relativistic optical jet
emission or by gravitational lensing of the continuum
but not the line emitting BLR. From the observed spectrum we obtain
an estimate of the enhancement needed in the optical continuum 
so that the emission
lines appear weak. This is done by calculating the equivalent width of 
Ly${\alpha}$ emission line from the observed spectrum of SDSS J1533$-$00 
and comparing it with 
the typical Ly${\alpha}$ emission line equivalent width observed in 
the composite spectrum of quasars (Francis et al.\ 1991). 
The equivalent width (EW) is defined as 
\begin{equation}
W_{obs} = \frac{F_{line}}{\mu F_{cont}}
\end{equation}
where $\mu$ is the magnification. The rest frame $W_{obs}$ is 9.9 \AA.  
When compared with the typical EW of about 80~\AA (Francis et al. 1991) 
observed in quasars, this gives a magnification of $\ge8$. Thus 
roughly an order of magnitude 
amplification in the continuum flux is needed to explain the observed weakness 
of the Ly$\alpha$ emission line if it is not due to a lack of emitting gas. 

\section{Discussion}
\subsection{Nature of optical variability of SDSS J1533$-$00}

The absence of strong emission lines in the spectrum of SDSS J1533$-$00 is 
similar to that of BL Lacs. It is also believed that the lack of emission 
lines in BL Lacs is due to the continuum being Doppler boosted. One of 
the defining characteristics of BL Lacs is that they show large amplitude
flux variability over the entire electro-magnetic spectrum.
Here, we discuss the observed variability properties of SDSS J1533$-$00 
and the known properties of the different kinds of AGNs derived based on 
long term optical monitoring to understand the nature of SDSS J1533$-$00.
The R band observations 
correspond to $\sim$1230~\AA ~in the QSO rest-frame.
A linear least squares fitting to our observations (see Eq.~1)
combined with two other epochs of observations of the SDSS team, gives a 
rate of change of $\sim$0.35 mag/yr in R band in the observed frame. 
This when transformed to the rest frame of
the quasar (taking into account the effects of time dilation; 
t$_{rest}$ = t$_{obs}$(1+{\it} z)) gives a rate of decline in the R band 
magnitude of $\sim$1.9 mag/yr. This rate of decline is an average, in
reality the decline could be more dramatic if there had been a flare in 
the intervening period.

Comparing SDSS and POSS measurements on a large sample of SDSS quasars,
de Vries et al. (2003) have reported that the long term quasar variability is 
consistent with a decaying intrinsic light-curve of the form, 
$1.1\times\exp(-t(yr)/2)$.  
Similar variability time-scales ($\sim$1 year) and amplitudes ($\sim$0.16 mag) 
have also been found by Trevese et al. (1994) from a 15 year monitoring of
35 quasars between  $0.6 < z < 3.1$. Giveon et al. (1999) have reported
an average rate of change in the intrinsic B band magnitude of 0.28 mag/year
using 7 years of monitoring data on 42 low-luminosity PG quasars sample with 
{\it z} $<$ 0.4.  For the same data, Cid Fernandes et al.\ (2000), using 
structure 
function analysis, report a variability amplitude and time-scale of 0.18 mag 
and 1.8 years respectively. Recently, using large database of variability on 
SDSS quasars, Ivezi{\'c} et al.\ (2004) report a variability amplitude and 
time-scale of 0.32 mag and 1 year, respectively. Thus, it appears that the normal 
population of QSOs seem to show an intrinsic rate of change 
of $\le 0.5$ mag/yr. In contrast, from the published lightcurves of blazars 
monitored over a 18 year time baseline (Webb et al. 1988), considering the 
long term trends, we notice that blazars on an average show a larger rate
of variability ($\ge$ 1 mag/yr). 

 From the above discussions, it is clear that the observed
R band variability of 1.9 mag/yr shown by SDSS J1533$-$00
is larger than that seen in typical QSO population.
As SDSS J1533$-$00 is highly luminous ($M_{1450\AA}$ = $-$26.6 mag),
X-ray and radio quiet and at high redshift, based on the existing correlations
(de Vries et al. 2004; Vanden Berk et al. 2004; Cid Fernandes et al. 1996;
Cristiani et al. 1996; Giveon et al. 1999), we expect it to show a lower rate of 
change of magnitude.  However, the observed high rate
of variability is consistent with that seen in  BL Lac
objects. Thus our observations are consistent with SDSS J1533$-$00
being a low luminosity AGN (with intrinsically weak emission lines) 
with its continuum being boosted by relativistic beaming. 
However, the continuum emission mechanism in the jet of this object
needs to be very different from typical BL Lac objects, since its broad-band
properties are so discordant.

Although the sparse long-term variability nature of
SDSS J1533$-$00 is similar to BL Lacs, intra-night monitoring observations
too are needed to confirm the presence of relativistic optical continuum
emitting jet in SDSS J1533$-$00. This is due to the fact that 
the observed large amplitude and frequent variability of 
BL Lacs on intra-night timescales (Stalin et al.\ 2004;
Gopal-Krishna et al.\ 2003) is now widely believed to be due to the 
inhomogeneities in the outflowing
relativistic jet. Such a study is planned in the coming observing season.

%It is believed that intra-night variability in AGNs with
%relativistic flows may be due to the inhomogeneities in the outflowing
%jet. Indeed, one of the defining characteristics of BL Lacs is that
%they show large amplitude and more frequent variability on intra-night
%time-scales compared to other classes of AGNs (Stalin et al.\ 2004; 
%Gopal-Krishna et al.\ 2003). Though the sparse long-term variability nature of 
%SDSS J1533$-$00 is similar to BL Lacs, intra-night monitoring observations 
%too are needed to confirm the presence of relativistic optical continuum
%emitting jet in SDSS J1533$-$00. 

\subsection{Is variability due to microlensing?}

In this section, we explore the possibility that the observed decline in the
light curve of SDSS J1533$-$00 is due to microlensing of the continuum source by an intervening 
object. Such a  model was proposed by Ostriker \& Vietri (1985) to explain
BL Lac objects (see also Gopal-Krishna and Subramanian 1991, for a 
relativistically moving source). Microlensing 
in the intervening galaxy is believed to be the cause of abnormal emission 
line ratios seen among different components in the multiply imaged systems 
such as Q2237+030 (Huchra et al.\ 1985), APM 08279+5255 (Lewis et al.\ 2002) 
and SDSS J1004+4112 (Richards et al.\ 2004). 
% THE NEXT SENTENCE IS REDUNDANT WITH THE FIRST IN THE PARA
%We consider here the possibility 
%if the variability observed in SDSS J1533$-$00 can be explained by 
%microlensing. 
As the QSO is not multiply imaged, the line of sight to the
QSO most probably samples the outer region of the intervening galaxy.
However, we note that the required magnification (i.e.,  $\ge 10$)
of the continuum for the emission lines to be invisible in SDSS J1533$-$00 
is much higher than that required in the case of multiply imaged
systems discussed above (i.e., $\sim$2).
\par
For the continuum source (and not the broad line emitting region) to be 
magnified significantly due to microlensing, the microlensing length scale, 
i.e., the Einstein radius, should be larger than the projected
radius of the continuum emitting region ($\le10^{14}$ cm) 
and smaller than the radius of the BLR (few $10^{17}$ cm). The 
Einstein radius ($R_E$) is defined as
\begin{equation}
R_E = \sqrt{\frac{4GM}{c^2} \frac{D_{ls}}{D_l D_s}}
\end{equation}
where $D_l$, $D_{ls}$ and $D_s$ are the lens, source-lens and 
source angular diameter distances, respectively, and $M$ is the 
mass of the lens. For a lens mid-way between the source and 
the observer (which corresponds to a lens redshift of 0.678), 
$R_E$ is $\sim$ 0.01 pc for 1 $M_{\odot}$ lens.
Such a lens will satisfy the requirements for microlensing described above.
We notice that to get the required high magnification the
impact parameter has to be less than 1.5$\times10^{15}$ cm
at the lens plane. The rate of change of magnification will depend 
upon the velocity of the star for a given impact parameter.

\begin{figure}
\hspace*{-0.5cm}\psfig{file=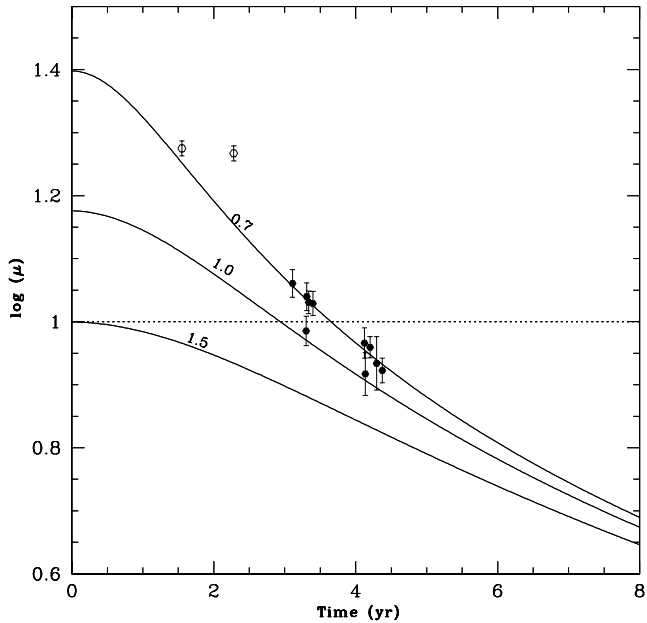,width=9cm,height=9cm}
\caption{Simulated microlensing lightcurves (log of magnification versus time) 
for three impact 
parameters. Our observations (filled circles) as well as those of 
Fan et al.\ (1999) (open circles) are overplotted on the modeled lensing 
lightcurves. The impact parameter in units of $10^{15}$ cm is given against 
each curve. The dotted horizontal line indicates a magnification of 10.}
\label{fig4}
\end{figure}

We obtain the expected light curve in the observer's frame assuming
the velocity of the lens to be 200 km s$^{-1}$ in the lens plane for
different impact parameters (using equation 20 of Refsdal 1964). 
The results are shown in Fig.~\ref{fig4}
where we plot the log of magnification as
a function of time. As we do not know the intrinsic flux and
the epoch at which SDSS J1533$-$00 was at its maximum
magnification, we shifted the observed points to match the predicted
curve with similar slope. From Fig.~4 it is clear that the sparsely sampled
observed light curve can match the predicted lensing curve for an
impact parameter of $\sim 7 \times10^{14}$ cm for an assumed solar mass
lens moving at 200 km s$^{-1}$. The required impact parameter will be slightly 
higher if we assume the velocity of the lensing star to be higher than
200 km s$^{-1}$. This 
simple exercise demonstrates that microlensing is a 
viable option in this case. Interestingly, the QSO is found to be
surrounded by excess density of galaxies, with the closest companion 
galaxy at a separation of only $\sim$3.5$^{\prime\prime}$ from the 
QSO (Hutchings 2003). We need closely sampled
multi-band light curves to confirm or refute this microlensing scenario.
Also, if microlensing is the correct option, then 
we now expect the Ly$\alpha$  
emission line to become more clearly visible in the spectrum of the QSO
as the enhanced continuum continues to decline.
Further photometric and spectroscopic observations should
thus constrain the microlensing hypothesis.

\subsection{How much gas is there in the BLR?}

\begin{table*}
\caption{Results of photoionization model computations. Here, N(H{\sc i}) = 
total hydrogen column density, $n_H$ = hydrogen density, 
R$_{BLR}$ = radius of the 
BLR, log U = logarithm of ionization parameter, L(Ly${\alpha}$) = luminosity of 
L(Ly${\alpha}$) emission line, $N_c$ = number of Ly${\alpha}$ emitting clouds
and $M_{BLR}$ = mass of the BLR }
\begin{tabular}{clccccr} \hline
 N(H{\sc i})   & ~~$n_H$     & R$_{BLR}$     & log U  & L(Ly${\alpha}$) & $N_c$  & $M_{BLR}$~~~~~ \\ 
 (cm$^{-2}$)   & (cm$^{-3}$) &   (pc)        &        &  (erg s$^{-1}$) &        &($M_{\odot}$)~~~~~ \\  \hline
 $10^{20}$  & $10^9$     &0.8 (0.17) &  $-$0.29 (~~0.07)  & 1.21 $\times$ $10^{28}$ (1.42 $\times$ $10^{28}$) & 1.9 $\times$ $10^{16}$ (1.6 $\times$ $10^{16}$) &  8.16 (6.96) \\
            & $10^{12}$  &0.8 (0.17) &  $-$3.29 ($-$2.93)  & 2.09 $\times$ $10^{24}$ (2.69 $\times$ $10^{24}$) & 1.1 $\times$ $10^{20}$ (8.4 $\times$ $10^{19}$) &   0.05 (00.04) \\ 
 $10^{21}$  & $10^9$     &0.8 (0.17) &  $-$0.29 (~~0.07)  & 8.36 $\times$ $10^{30}$ (8.07 $\times$ $10^{30}$) & 2.7 $\times$ $10^{13}$ (2.8 $\times$ $10^{13}$) &  11.83 (12.25) \\
            & $10^{12}$  &0.8 (0.17) &  $-$3.29 ($-$2.93)  & 2.91 $\times$ $10^{26}$ (6.26 $\times$ $10^{26}$) & 7.7 $\times$ $10^{17}$ (3.6 $\times$ $10^{17}$) &   0.34 (00.16) \\ 
 $10^{22}$  & $10^9$     &0.8 (0.17) &  $-$0.29 (~~0.07)  & 7.76 $\times$ $10^{33}$ (7.00 $\times$ $10^{33}$) & 2.9 $\times$ $10^{10}$ (3.2 $\times$ $10^{10}$) &  12.75 (14.13) \\
            & $10^{12}$  &0.8 (0.17) &  $-$3.29 ($-$2.93)  & 2.92 $\times$ $10^{28}$ (6.30 $\times$ $10^{28}$) & 7.7 $\times$ $10^{15}$ (3.6 $\times$ $10^{15}$) &   3.38 (01.57) \\ 
 $10^{23}$  & $10^9$     &0.8 (0.17) &  $-$0.29 (~~0.07)  & 6.04 $\times$ $10^{36}$ (6.21 $\times$ $10^{36}$) & 3.7 $\times$ $10^{07}$ (3.6 $\times$ $10^{07}$) &  16.38 (15.92) \\
            & $10^{12}$  &0.8 (0.17) &  $-$3.29 ($-$2.93)  & 2.92 $\times$ $10^{30}$ (6.30 $\times$ $10^{30}$) & 7.7 $\times$ $10^{13}$ (3.6 $\times$ $10^{13}$) &  33.83 (15.70) \\ \hline
\end{tabular}
\end{table*}

\par
The upper limit we derive on the Ly$\alpha$ line flux (see Sect. 2.2) can be
used to obtain a limit on the mass of the line emitting gas in the
BLR. If BLR gas emits very efficiently then the mass of the
BLR ($M_{BLR}$) is given by
\begin{equation}
M_{BLR} = 5.1 (10^{11}/n_e) (L_{{\rm Ly}\alpha}/10^{45})~M_{\odot}
\end{equation}
For the upper limit of Ly$\alpha$ luminosity of SDSS 1533$-$00,
we get $M_{BLR}\le1.1~M_\odot$ for $n_e = 10^{11}$  cm$^{-3}$.
As pointed out by Baldwin et al.\ (2003) the mass estimated using
the above equation will be the minimum mass in the BLR. We
obtain a realistic upper limit to $M_{BLR}$ using the 
photoionization code CLOUDY (Ferland et al. 1998).
\par
The ionization state  of the line emitting gas is quantified using
a dimensionless ionization parameter, $U$. This is
defined as,
\begin{equation}
U = \frac{L_{912}}{4\pi R_{BLR}^2 h \nu c n}
\end{equation}
where $n$, $L_{912}$  and $R_{BLR}$ are, respectively, the number density of 
the gas particles, the Lyman continuum luminosity, and the radius of the
BLR around the central engine. We obtain
$L_{912}~=~2.7\times10^{46}$ erg s$^{-1}$ using the observed
flux above the Ly$\alpha$ emission line and a spectral
index, $\alpha_o = -0.8$ (see Sect. 2.2). For a standard AGN, 
$R_{BLR}$ can be estimated from the rest-frame luminosity at 
5100~\AA (i.e., ${L_{5100}}$) using
\begin{equation}
R_{BLR} = 27.4 \left( \frac{L_{5100}}{10^{44} {\rm erg s}^{-1}}\right)^{0.68}{\rm
 light~days},
\end{equation}
(see Corbett et al.\ 2003 and references therein). The above relationship is
established using reverberation mapping studies. Assuming such a relationship 
holds good even for SDSS~J1533$-$00, we estimate the radius of the BLR to be 
0.8 pc. Thus we have a constraint on $U$ for an assumed value of $n$. For example, 
when we consider $n = 10^9$ cm$^{-3}$ the consistent value of log$U$ is $-$0.29.

We then estimated the Ly$\alpha$ emission line luminosity from a single cloud
($L({\rm Ly}{\alpha})^{cal}$) for a given n, U, total hydrogen column density 
N(H{\sc i}) and ionizing spectrum of the QSO using the photoionization code 
CLOUDY (Ferland et al. 1998). This together with the observed upper limit on 
the Ly$\alpha$ emission line flux ($ F(Ly{\alpha})^{obs}$) is used to 
get the number of BLR clouds,
\begin{equation}
N_c = \frac{4 \pi d_l^2 F({\rm Ly}{\alpha})^{obs}} {L({\rm Ly}{\alpha})^{cal}}.
\end{equation}
Here, for simplicity we assume all the BLR clouds to be identical.
In addition if we assume the clouds to be spherical we can get
the mass of the Ly$\alpha$ emitting clouds in the BLR. 
The total mass of the BLR contained in $N_c$ clouds
is calculated using
\begin{equation}
M_{BLR} = N_c \times {4\over3}\pi r^3 \rho
\end{equation}
where $\rho$ and  $r$ are the total hydrogen mass density and 
radius of the cloud ($r$ = N(H{\sc i})/2n), respectively.

The ionizing continuum used in the model calculations is a combination of a 
UV bump (assumed to be a black-body with a temperature of 10$^6$ K) 
and an X-ray power law (Georgantopoulos et al. 2004) of the form
$f_{\nu} \propto \nu^{-2.0}$. The UV and the X-ray continuum slopes were
combined using an UV to X-ray logarithmic spectral slope of 
$\alpha_{ox}$ = $-$2.0. This value of $\alpha_{ox}$ is consistent
with the upper limit derived by Vignali et al.\ (2003). 

For this assumed incident ionizing continuum shape and solar chemical 
abundances, we carried out grids of photoionization model computations for 
various values of N(H{\sc i}) and $n$. The chosen range in N(H{\sc i}) considers
BLR to range from optically thin to optically thick. The ranges of 
other chosen parameters are also consistent with those frequently used in 
BLR modeling (see Korista et al.\ 1997). The results of these model
calculations are summarized in Table 2. It is found that the
calculated values of the $Ly{\alpha}$ emission line luminosity are weakly dependent
on the input values of $\alpha_{ox}$. We also notice that changing the 
temperature of the UV bump from $10^5$ K to $10^6$ K changes the luminosity by 
less than a factor of 2. Assuming the observed continuum is enhanced 
by a factor of 10 (due to either Doppler boosting or gravitational
lensing), we also carried out photoionization model computations
by de-magnifying the measured continuum luminosity by a factor of 10.
The results are shown within brackets in Table 2.

Clearly our model calculations are consistent with $M_{BLR} < 50~M_\odot$
in the case of SDSS J1533$-$00. This is 1 to 2 orders of magnitude
less than that derived for the standard high luminosity QSOs
(Baldwin et al.\ 2003). However, the estimated upper limit is
consistent with BLR masses computed for low luminosity Seyfert galaxies 
(see Peterson 2004).

%%%%%%%%%%%%
\section{Conclusion}
\label{sec:conclusion}
We have presented optical photometric monitoring of the peculiar quasar 
SDSS J1533$-$00 for a duration spanning about 500 days during which the object 
has varied by about 0.4 mag. These observations, when coupled with two other
epochs of observations available in the literature, indicate that the quasar 
gradually faded by 0.9 magnitude during the period June 1998 to March 2001. 
This transforms to a variability of $\sim$1.9 mag/yr in the
rest frame of the quasar. Such a large amplitude of variability is similar to 
the long term variability nature of BL Lacs.
Nevertheless, the lack of X-ray and radio emission and optical 
polarization suggests that the continuum emission in the jet needs to be
very different from BL Lacs. Available photometric data on the source could 
not rule out microlensing as the cause of the variability as well as the observed 
lack of emission lines. Further monitoring observations are needed to 
constrain the microlensing scenario. 

Photoionization model calculations
show the BLR mass to be consistent with low luminosity Seyferts (Peterson 2004),
but $\sim$2 orders of magnitude lower compared to those expected for high luminosity
quasars (Baldwin et al.\ 2003). It is also possible that SDSS J1533$-$00 
belongs to an unknown population of highly luminous AGNs without
a BLR. This would be in line with the ``naked" AGNs discussed by Hawkins (2004). 
Furthermore, if the observed lack of emission lines in the discovery
spectrum of SDSS J1533$-$00 (Fan et al. 1999) is due to the 
continuum being amplified either due to Doppler boosting 
or to gravitational lensing, then we might expect to see emission lines 
emerge as
our observations show the quasar to be in the declining phase.
%%%%%%%%%%%%
Further photometric and spectroscopic observations
could clarify the peculiar nature of SDSS J1533$-$00.

There now exist quite a few other objects with 
featureless optical spectra resembling BL Lacs but which  lack the
significant radio and X-ray emission found in BL Lacs (Anderson et al. 2001; Hall et al. 2004; 
Londish et al. 2002). 
Whether SDSS J1533$-$00, along with such other
objects, fit into the orientation based unification scheme, or whether they 
instead belong to a hitherto 
unrecognized population of radio-quiet BL Lacs or lineless radio-quiet quasars 
remains an  open question.

%%%%%%%%%%%%%%%%%%%%%%%%%%%
\section*{Acknowledgements}
%%%%%%%%%%%%%%%%%%%%%%%%%%%
It is pleasure to thank Dr. X. Fan for providing us the Keck II LRIS spectrum 
of SDSS J153259.96$-$003944.1 and the referee for useful comments. We 
also thank Profs. P. J. Wiita, Gopal-Krishna and Ram Sagar for critical 
comments and suggestions as well as Prof. K. Subramanian for useful discussions.
CSS thanks the Virtual Observatory-India project 
for financial support for this work.

\end{document}